\documentclass[aps,prl,showpacs,twocolumn,superscriptaddress,preprintnumbers,amsmath,amssymb]{revtex4}
\usepackage{graphicx} % Include figure files
\usepackage{dcolumn}  % Align table columns on decimal point
\usepackage{subfigure}

\graphicspath{{ps}}

\renewcommand{\arraystretch}{1.1}

\newcommand{\pip}{\pi^+}
\newcommand{\pim}{\pi^-}

\newcommand{\km}{K^-}

\newcommand{\kl}{K^0_L}

\newcommand{\lamc}{\Lambda_c^+}
\newcommand{\sigc}{\Sigma_c(2455)}
\newcommand{\scst}{\Sigma_c(2520)}
\newcommand{\lcf}{\Lambda_c(2880)^+}
\newcommand{\lcs}{\Lambda_c(2940)^+}
\newcommand{\lct}{\Lambda_c(2765)^+}

\newcommand{\mev}{\mathrm{MeV}}
\newcommand{\gev}{\mathrm{GeV}}

\newcommand{\mevm}{\mathrm{MeV}/c^2}
\newcommand{\gevm}{\mathrm{GeV}/c^2}
\newcommand{\costh}{\cos \theta}

\newcommand{\rone}{0.404\pm0.021\pm0.014}
\newcommand{\rtwo}{0.091\pm0.025\pm0.010}
\newcommand{\ronetwo}{0.225\pm0.062\pm0.025}

\newcommand{\siglcs}{7.2}
\newcommand{\siglcssy}{6.2}

\newcommand{\sigsta}{3.7}

\newcommand{\sigfo}{5.5}
\newcommand{\sigfosy}{5.5}

\newcommand{\sigft}{4.8}
\newcommand{\sigftsy}{4.5}

\begin{document}

\title{\boldmath Experimental Constraints on the Spin and Parity of
the $\lcf$}

%\author{The Belle Collaboration}
\date{\today}

\begin{abstract}
\noindent
We report the results of several studies of the $\lamc\pip\pim X$
final state in continuum $e^+e^-$ annihilation data collected by the
Belle detector.
An analysis of angular distributions in $\lcf\to\sigc^{0,++}\pi^{+,-}$
decays strongly favors a $\lcf$ spin assignment of $\frac{5}{2}$ over
$\frac{3}{2}$ or $\frac{1}{2}$.
We find evidence for $\lcf\to\scst^{0,++}\pi^{+,-}$ decay and measure
the ratio of $\lcf$ partial widths
$\frac{\Gamma(\scst\pi)}{\Gamma(\sigc\pi)}=\ronetwo$. This value
favors the $\lcf$ spin-parity assignment of $\frac{5}{2}^+$ over
$\frac{5}{2}^-$.
We also report the first observation of $\lcs\to\sigc^{0,++}\pi^{+,-}$
decay and measure $\lcf$ and $\lcs$ parameters.
These studies are based on a $553\,\mathrm{fb}^{-1}$ data sample
collected at or near the $\Upsilon(4S)$ resonance, at the KEKB
collider. 
\end{abstract}


%%% Paper:    Lambda_c(2880)+
%%% Journal:  Physical Review Letters
%%% Contacts: R. Mizuk (mizuk@itep.ru)
%%% Non-responding authors or those who said NO are commented out.
%%% ====================================================================
%%% Click the RELOAD button on your web browser to see the updated file.
%%% ====================================================================
%%% Use \input{author} to insert this material into your latex file.
%%%%% Force institutions to appear in alphabetical order when typeset.
\affiliation{Budker Institute of Nuclear Physics, Novosibirsk}
%%%\affiliation{Chiba University, Chiba}
\affiliation{Chonnam National University, Kwangju}
\affiliation{University of Cincinnati, Cincinnati, Ohio 45221}
\affiliation{Department of Physics, Fu Jen Catholic University, Taipei}
%%%\affiliation{Justus-Liebig-Universit\"at Gie\ss{}en, Gie\ss{}en}
\affiliation{The Graduate University for Advanced Studies, Hayama, Japan}
%%%\affiliation{Gyeongsang National University, Chinju}
\affiliation{University of Hawaii, Honolulu, Hawaii 96822}
\affiliation{High Energy Accelerator Research Organization (KEK), Tsukuba}
%%%\affiliation{Hiroshima Institute of Technology, Hiroshima}
\affiliation{University of Illinois at Urbana-Champaign, Urbana, Illinois 61801}
\affiliation{Institute of High Energy Physics, Chinese Academy of Sciences, Beijing}
\affiliation{Institute of High Energy Physics, Vienna}
\affiliation{Institute of High Energy Physics, Protvino}
\affiliation{Institute for Theoretical and Experimental Physics, Moscow}
\affiliation{J. Stefan Institute, Ljubljana}
\affiliation{Kanagawa University, Yokohama}
\affiliation{Korea University, Seoul}
%%%\affiliation{Kyoto University, Kyoto}
\affiliation{Kyungpook National University, Taegu}
\affiliation{Swiss Federal Institute of Technology of Lausanne, EPFL, Lausanne}
\affiliation{University of Ljubljana, Ljubljana}
\affiliation{University of Maribor, Maribor}
\affiliation{University of Melbourne, Victoria}
\affiliation{Nagoya University, Nagoya}
\affiliation{Nara Women's University, Nara}
\affiliation{National Central University, Chung-li}
\affiliation{National United University, Miao Li}
\affiliation{Department of Physics, National Taiwan University, Taipei}
\affiliation{H. Niewodniczanski Institute of Nuclear Physics, Krakow}
\affiliation{Nippon Dental University, Niigata}
\affiliation{Niigata University, Niigata}
\affiliation{University of Nova Gorica, Nova Gorica}
\affiliation{Osaka City University, Osaka}
\affiliation{Osaka University, Osaka}
\affiliation{Panjab University, Chandigarh}
\affiliation{Peking University, Beijing}
%%%\affiliation{University of Pittsburgh, Pittsburgh, Pennsylvania 15260}
\affiliation{Princeton University, Princeton, New Jersey 08544}
\affiliation{RIKEN BNL Research Center, Upton, New York 11973}
%%%\affiliation{Saga University, Saga}
\affiliation{University of Science and Technology of China, Hefei}
\affiliation{Seoul National University, Seoul}
\affiliation{Shinshu University, Nagano}
\affiliation{Sungkyunkwan University, Suwon}
\affiliation{University of Sydney, Sydney NSW}
\affiliation{Tata Institute of Fundamental Research, Bombay}
\affiliation{Toho University, Funabashi}
\affiliation{Tohoku Gakuin University, Tagajo}
\affiliation{Tohoku University, Sendai}
\affiliation{Department of Physics, University of Tokyo, Tokyo}
\affiliation{Tokyo Institute of Technology, Tokyo}
\affiliation{Tokyo Metropolitan University, Tokyo}
%%%\affiliation{Tokyo University of Agriculture and Technology, Tokyo}
%%%\affiliation{Toyama National College of Maritime Technology, Toyama}
%%%\affiliation{University of Tsukuba, Tsukuba}
\affiliation{Virginia Polytechnic Institute and State University, Blacksburg, Virginia 24061}
\affiliation{Yonsei University, Seoul}
  \author{R.~Mizuk}\affiliation{Institute for Theoretical and Experimental Physics, Moscow} % ITEP
  \author{K.~Abe}\affiliation{High Energy Accelerator Research Organization (KEK), Tsukuba} % KEK
% \author{K.~Abe}\affiliation{Tohoku Gakuin University, Tagajo} % TohokuGakuin
% \author{N.~Abe}\affiliation{Tokyo Institute of Technology, Tokyo} % TIT
  \author{I.~Adachi}\affiliation{High Energy Accelerator Research Organization (KEK), Tsukuba} % KEK
  \author{H.~Aihara}\affiliation{Department of Physics, University of Tokyo, Tokyo} % Tokyo
  \author{D.~Anipko}\affiliation{Budker Institute of Nuclear Physics, Novosibirsk} % BINP
% \author{K.~Aoki}\affiliation{Nagoya University, Nagoya} % Nagoya
% \author{K.~Arinstein}\affiliation{Budker Institute of Nuclear Physics, Novosibirsk} % BINP
% \author{Y.~Asano}\affiliation{University of Tsukuba, Tsukuba} % Tsukuba
% \author{T.~Aso}\affiliation{Toyama National College of Maritime Technology, Toyama} % Toyama
  \author{V.~Aulchenko}\affiliation{Budker Institute of Nuclear Physics, Novosibirsk} % BINP
  \author{T.~Aushev}\affiliation{Swiss Federal Institute of Technology of Lausanne, EPFL, Lausanne}\affiliation{Institute for Theoretical and Experimental Physics, Moscow} % ITEP
% \author{T.~Aziz}\affiliation{Tata Institute of Fundamental Research, Bombay} % Tata
% \author{S.~Bahinipati}\affiliation{University of Cincinnati, Cincinnati, Ohio 45221} % Cincinnati
  \author{A.~M.~Bakich}\affiliation{University of Sydney, Sydney NSW} % Sydney
  \author{V.~Balagura}\affiliation{Institute for Theoretical and Experimental Physics, Moscow} % ITEP
% \author{Y.~Ban}\affiliation{Peking University, Beijing} % Peking
% \author{S.~Banerjee}\affiliation{Tata Institute of Fundamental Research, Bombay} % Tata
  \author{E.~Barberio}\affiliation{University of Melbourne, Victoria} % Melbourne
% \author{M.~Barbero}\affiliation{University of Hawaii, Honolulu, Hawaii 96822} % Hawaii
% ...
% Ljubljana
% \author{D.~Z\"urcher}\affiliation{Swiss Federal Institute of Technology of Lausanne, EPFL, Lausanne} % Lausanne
\collaboration{The Belle Collaboration}

\maketitle

{\renewcommand{\thefootnote}{\fnsymbol{footnote}}}
\setcounter{footnote}{0}

Charmed baryon spectroscopy provides an excellent laboratory to study the
dynamics of a light diquark in the environment of a heavy quark,
allowing the predictions of different theoretical approaches to be
tested~\cite{capstick_isgur,hqs,chi_sm,relat_qm}.
There are twelve experimentally observed charmed baryons for which the
spins and parities are assigned~\cite{pdg,babar_omcst}. They include
ground states, spin excitations and lowest orbital excitations.
Except for the $\lamc$, the $J^P$ quantum numbers for these states
have not been determined experimentally but are assigned based on the
quark model predictions for their masses. There are also six charmed
baryons, recently observed at the CLEO~\cite{cleo_lamc2880},
Belle~\cite{sigc2800,xic} and BaBar~\cite{babar_lamc2940} experiments,
for which the spins and parities are not well constrained. The new
states are in a mass region where the quark model predicts many levels
with small spacing, which makes the $J^P$ assignment difficult.
In this Letter we investigate possible spin and parity values of one
such state, the $\lcf$ baryon~\cite{cleo_lamc2880,babar_lamc2940},
by studying the resonant structure of $\lcf\to\lamc\pip\pim$ decays
and performing an angular analysis of $\lcf\to\sigc^{0,++}\pi^{+,-}$
decays.
We also report the first observation of $\lcs\to\sigc^{0,++}\pi^{+,-}$
decay and measure $\lcf$ and $\lcs$ parameters.

We use a $553\,\mathrm{fb}^{-1}$ data sample collected with the Belle
detector at or $60\,\mev$ below the $\Upsilon(4S)$ resonance, at the
KEKB asymmetric-energy ($3.5\,\gev$ on $8.0\,\gev$) $e^+e^-$
collider~\cite{KEKB}.
The Belle detector~\cite{BELLE_DETECTOR} is a large-solid-angle
magnetic spectrometer that consists of a silicon vertex detector, a
50-layer cylindrical drift chamber, an array of aerogel threshold
Cherenkov counters, a barrel-like array of time-of-flight
scintillation counters, and an array of CsI(Tl) crystals located
inside a superconducting solenoidal coil that produces a 1.5T magnetic
field. An iron flux return located outside the coil is instrumented to
detect muons and $\kl$ mesons.
We use a GEANT based Monte-Carlo (MC) simulation~\cite{geant} to model
the response of the detector and to determine its acceptance. Signal
MC events are generated with experimental run dependence in proportion
to the relative luminosities of different running periods.

$\lamc$ baryons are reconstructed using the $p K^-\pi^+$ decay mode
(the inclusion of charge conjugate modes is implied throughout this
Letter). To select proton, charged kaon and pion candidates we use the
same track quality and particle identification criteria as for
observation of the $\Sigma_c(2800)$ isotriplet~\cite{sigc2800}.
The invariant mass of the $p\km\pip$ combinations is required to be
within $\pm8\,\mevm$ ($1.6\sigma$) of the $\lamc$ mass value, recently
measured by BaBar~\cite{babar_mlamc}.
To improve the accuracy of the $\lamc$ momentum measurement we perform
a mass constrained fit to the $p\km\pip$ vertex.
We combine $\lamc$ candidates with the remaining $\pip\pim$ candidates
in the event. 
To reduce the combinatorial background we impose a requirement on the
scaled momentum of the $\lamc\pip\pim$ combination
$x_p \equiv p^\ast/\sqrt{E^{\ast2}_{\rm beam}-M^2}>0.7$, where
$p^\ast$ is the momentum and $M$ is the invariant mass of the
combination, $E^\ast_{\rm beam}$ is the beam energy, all variables
being measured in the center-of-mass frame. To improve the
$M(\lamc\pip\pim)$ resolution we perform an interaction point
constrained fit to the $\lamc\pip\pim$ vertex.

\begin{figure}[ptbh]
\centering
\begin{picture}(550,170)
\put(20,80){\rotatebox{90}{${\rm Events}\;/\; 2.5\,\mevm$}} 
\put(80,0){$M(\lamc\pip\pim),\,\gevm$} 
\put(10,-20){\includegraphics[width=8cm]{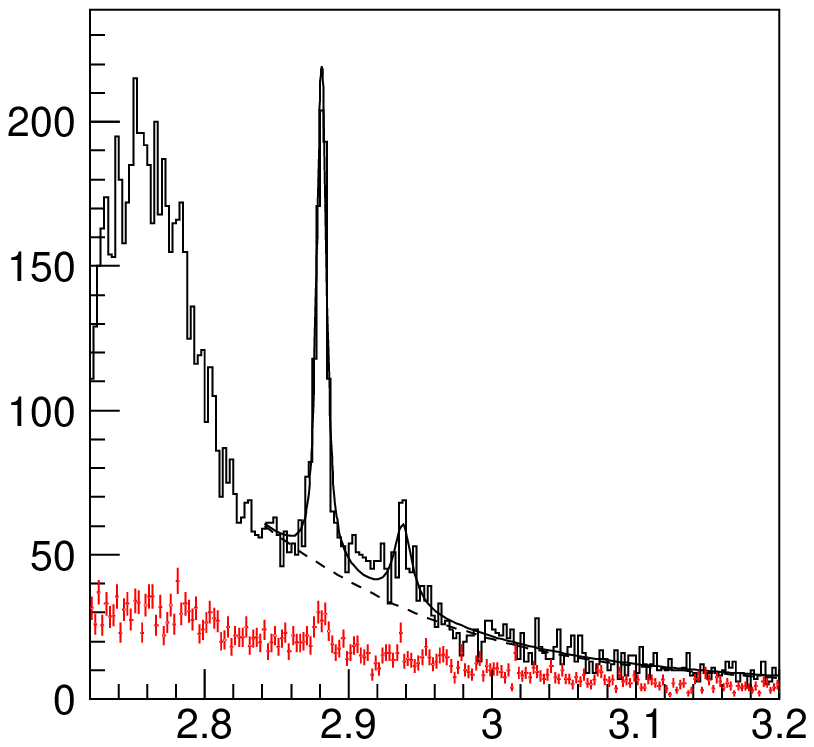}}
\end{picture}
\caption{The invariant mass of the $\lamc\pip\pim$ combinations for
  the $\sigc$ signal region (histogram) and scaled sidebands (dots
  with error bars). The fit result (solid curve) and its combinatorial
  component (dashed curve) are also presented. }
\label{fig1}
\end{figure}

To measure the $\lcf$ mass and width we apply an additional
requirement that either $M(\lamc\pim)$ or $M(\lamc\pip)$ be in the
$\sigc$ signal region defined as $2450\,\mevm<M<2458\,\mevm$. 
Whereas 35\% of signal events pass this cut, only 12\% of background
events do so.
From MC simulation we find that the mass resolution for the
$\lcf\to\sigc^{0,++}\pi^{+,-}$ decays depends strongly on the decay
angle $\theta$, defined as the angle between the pion momentum in the
$\lcf$ rest frame and the boost direction of the $\lcf$.
To assure good resolution for the $\lcf$ mass and width measurement we
require \mbox{$\costh>0$}. This requirement also helps to suppress
combinatorial background.
The resulting $M(\lamc\pip\pim)$ distribution is shown in
Fig.~\ref{fig1}. 
One can see clear peaks from the $\Lambda_c(2765)^+$ and $\lcf$.  
A peak in the region $M=2940\,\mevm$ is associated with the $\lcs$
baryon recently observed in the $D^0p$ final state by
BaBar~\cite{babar_lamc2940}.
Scaled $\sigc$ sidebands, which are also shown in Fig.~\ref{fig1}, are
featureless in the region of the $\lcs$.  The $\sigc$ sidebands are
defined as $2438\,\mevm<M(\lamc\pi)<2446\,\mevm$ and
$2462\,\mevm<M(\lamc\pi)<2470\,\mevm$. 

We perform a binned likelihood fit to the $\lamc\pip\pim$ mass
spectrum of Fig.~\ref{fig1} to extract the parameters and yields of
the $\lcf$ and $\lcs$.  The fitting function is a sum of three
components: $\lcf$ signal, $\lcs$ signal and combinatorial background
functions. 
As shown below, the favored spin-parity assignment for the $\lcf$ is
$\frac{5}{2}^+$, therefore the $\lcf$ signal is parameterized by an
F-wave Breit-Wigner function convolved with the detector resolution
function, determined from MC ($\sigma=2.2\,\mevm$). The $\lcs$ signal
is an S-wave Breit-Wigner function convolved with the detector
resolution function ($\sigma=2.4\,\mevm$).
The background is parameterized by a third-order polynomial. The fit
is shown in Fig.~\ref{fig1}, and the results are summarized in
Table~\ref{tab_94_fit}.
\begin{table}[htb]
\caption{Signal yield, mass and width for the $\lcf$ and $\lcs$. The
  first uncertainty is statistical, the second one systematic.}
\label{tab_94_fit}
\renewcommand{\arraystretch}{1.2}
\begin{ruledtabular}
\begin{tabular}{lccc}
State & Yield & $M,\;\mevm$ & $\Gamma,\;\mev$ \\
\hline
$\lcf$ & $690\pm50$ & $2881.2\pm0.2\pm0.4$ & $5.8\pm0.7\pm1.1$ \\
$\lcs$ & $220^{+80}_{-60}$ & $2938.0\pm1.3^{+2.0}_{-4.0}$ & $13^{+8}_{-5}{^{+27}_{-\phantom{2}7}}$ \\
\end{tabular} 
\end{ruledtabular}
\end{table}
The signal yield is defined as the integral of the Breit-Wigner
function over a $\pm2.5\Gamma$ interval.
The normalized $\chi^2$ of the fit is $\chi^2/d.o.f.=132.2/134$.
If the $\lcs$ signal is removed from the fit, the double log
likelihood changes by 59.8, which corresponds (for 3 degrees of
freedom) to a signal significance of $\siglcs$ standard deviations.

To estimate the systematic uncertainty on the results of the fit we
vary the background parameterization, using a fourth-order polynomial
and the inverse of a third-order polynomial.
We include the $\lct$ signal region into the fit interval,
parameterizing the $\lct$ signal by an S-wave Breit-Wigner
function. The $\lct$ mass and width determined from the fit are
$M=(2761\pm1)\,\mevm$ and $\Gamma=(73\pm5)\,\mev$.
We vary the selection requirements;
we take into account the uncertainty in the $\lamc$ mass of
$\pm0.14\,\mevm$~\cite{babar_mlamc}, the mass scale uncertainty of
$^{+0.19}_{-0.21}\,\mevm$~\cite{xic_mass} and the uncertainty in the
detector resolution of $\pm10\%$ as estimated by comparison of the
inclusive $\lamc\to pK^-\pi^+$ signal in data and MC.
In the region between the $\lcf$ and $\lcs$ signals the fit is
systematically below the data points, which might be due to a presence
of an additional resonance or due to interference.
We take into account these possibilities as a systematic uncertainty.
In each case we consider the largest positive and negative variation
in the $\lcf$ and $\lcs$ parameters to be the systematic uncertainty
from this source; each term is then added in quadrature to give the
total systematic uncertainty, quoted in Table~\ref{tab_94_fit}.
The main sources of the systematic uncertainty are a possible
contribution of the $\lct$ tail into the fit region (the shape of the
tail is not well constrained) and the excess of events between the
$\lcf$ and $\lcs$ signals.
None of the variations in the analysis alters the $\lcs$ signal
significance to less than $\siglcssy$ standard deviations.

\begin{figure}[ptbh]
\centering
\begin{picture}(550,170)
\put(20,87){\rotatebox{90}{${\rm Events}\;/\; 2\,\mevm$}} 
\put(90,0){$M(\lamc\pi^{\pm}),\,\gevm$} 
\put(10,-20){\includegraphics[width=8cm]{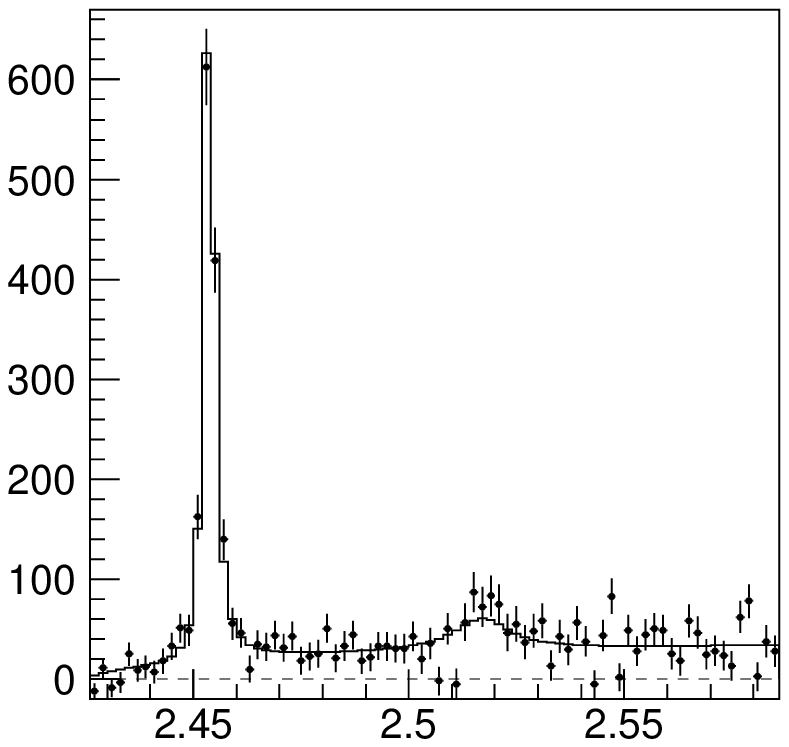}}
\end{picture}
\caption{The $\lcf$ yield as a function of $M(\lamc\pi^{\pm})$. The
  histogram represents the result of the fit.}
\label{fig2}
\end{figure}

For further analysis, we remove the $\costh>0$ requirement. 
To study the resonant structure of the $\lcf\to\lamc\pip\pim$ decays
we fit the $\lamc\pip\pim$ mass spectrum in $M(\lamc\pi^{\pm})$
bins. By isospin symmetry, we expect equally many decays to proceed
via a doubly charged $\sigc$ ($\scst$) as via a neutral one.
Since the corresponding doubly charged and neutral channels are
kinematically separated in phase space, we combine the
$M(\lamc\pip\pim)$ distributions for $M(\lamc\pim)$ and $M(\lamc\pip)$
bins.
To fit the $\lamc\pip\pim$ mass spectra we use the same fit function
as described above.
The $\lcf$ and $\lcs$ parameters are fixed to the values in
Table~\ref{tab_94_fit}.
The $\lcf$ yield as a function of $M(\lamc\pi^{\pm})$ is
shown in Fig.~\ref{fig2}.
We find a clear signal for the $\sigc$ and an excess of events in
the region of the $\scst$.
We perform a $\chi^2$ fit to the $\lamc\pi^{\pm}$ mass spectrum of
Fig.~\ref{fig2} to extract the yields of the $\sigc$ and $\scst$. The
fitting function is a sum of three components: $\sigc$ signal, $\scst$
signal and a non-resonant contribution.
The $\sigc$ and $\scst$ signals are parameterized by a P-wave
Breit-Wigner function convolved with the detector resolution
functions, determined from MC ($\sigma=0.9\,\mevm$ for the $\sigc$ and
$\sigma=1.5\,\mevm$ for the $\scst$).
The mass and width of the $\sigc$ are floated, while the mass and
width of the $\scst$ are fixed to the world average values~\cite{pdg}.
The shape of the non-resonant contribution is determined from MC
assuming a uniform distribution of the signal over phase space. 
The fit is shown in Fig.~\ref{fig2}. 
We find the ratios of $\lcf$ partial widths
$\frac{\Gamma(\sigc\pi^{\pm})}{\Gamma(\lamc\pip\pim)}=\rone$,
$\frac{\Gamma(\scst\pi^{\pm})}{\Gamma(\lamc\pip\pim)}=\rtwo$ and
$\frac{\Gamma(\scst\pi^{\pm})}{\Gamma(\sigc\pi^{\pm})}=\ronetwo$,
where the uncertainties are statistical and systematic, respectively.
The $\sigc$ parameters determined from the fit $M=(2453.7\pm0.1)\,\mevm$
and $\Gamma=(2.0\pm0.2)\,\mev$ are consistent with the world average
values~\cite{pdg}.
The normalized $\chi^2$ of the fit is $\chi^2/d.o.f.=106.6/75$.
The significance of the $\scst$ signal is $\sigsta$ standard deviations. 

To estimate the systematic uncertainties on the ratios of $\lcf$
partial widths we vary the $\lcf$ parameters, fit interval and
background parameterization in the fit to the $M(\lamc\pip\pim)$
spectrum; we vary the $\scst$ parameters; we allow the shape of the
non-resonant contribution to float in the fit, parameterizing it with a
second-order polynomial multiplied by a threshold function or by a
third-order polynomial; we take into account the uncertainty in the
detector resolution and in the reconstruction efficiency.
None of the variations reduces the significance of the $\scst$ signal
below three standard deviations.

\begin{figure}[ptbh]
\centering
\begin{picture}(550,320)
\put(18,260){\rotatebox{90}{${\rm Events}\;/\; 0.2$}} 
\put(18,85){\rotatebox{90}{${\rm Events}\;/\; (\pi/10)$}} 
\put(120,163){$\cos\theta$} 
\put(120,2){$\phi$, rad} 
\put(10,-20){\includegraphics[width=8cm]{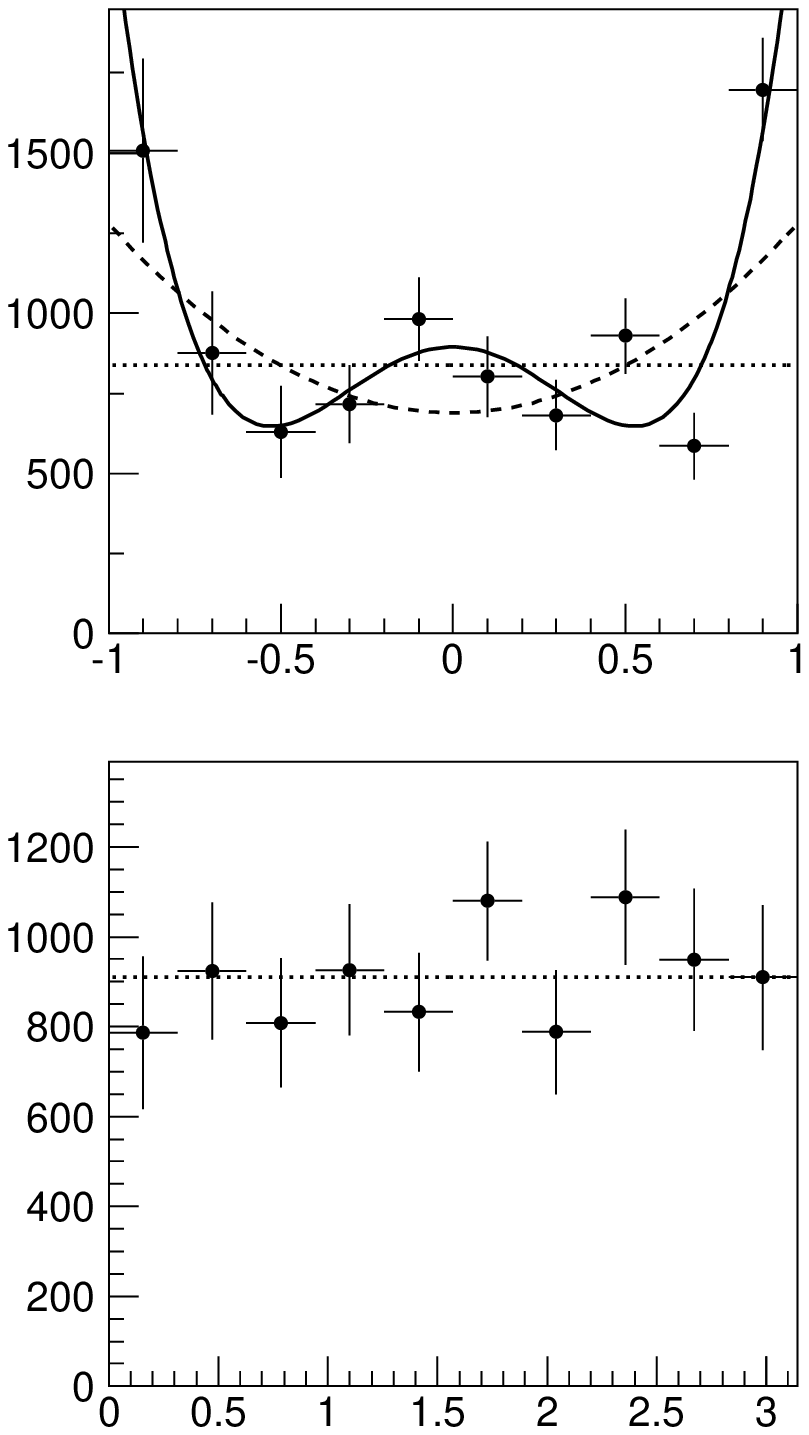}}
\end{picture}
\caption{The yield of $\lcf\to\sigc^{0,++}\pi^{+,-}$ decays as a
  function of $\costh$ and $\phi$. The fits are described in the
  text.}
\label{fig3}
\end{figure}

To perform angular analysis of $\lcf\to\sigc^{0,++}\pi^{+,-}$ decays
we fit the $\lamc\pip\pim$ spectrum in $\costh$ and $\phi$ bins
for the $\sigc$ signal region and sidebands. 
Here, $\phi$ is the angle between the $e^+e^-\to\lcf X$ reaction plane
and the plane defined by the pion momentum and the $\lcf$ boost
direction in the rest frame of the $\lcf$.
Figure~\ref{fig3} shows the yield of $\lcf$ as a function of $\costh$
and $\phi$, after $\sigc$ sideband subtraction (to account for
nonresonant $\lamc\pip\pim$ decays) and efficiency correction.

The parameterization of $\lcf\to\sigc\pi$ decay angular distributions
depends on the spin of the $\lcf$. For the spin $\frac{1}{2}$
hypothesis both $\cos\theta$ and $\phi$ distributions are expected to
be uniform~\cite{pilkuhn}. $\chi^2$ fits to a constant are shown in
Fig.~\ref{fig3} by a dotted line.
The agreement is good for $\phi$: $\chi^2/d.o.f.=5.3/9$, but poor for
$\costh$: $\chi^2/d.o.f.=46.7/9$.

The angular distribution for the spin $\frac{3}{2}$ hypothesis 
is~\cite{pilkuhn} 
\begin{align*}
W_{3/2}=\frac{3}{4\pi}[\rho_{33}\sin^2\theta+
\rho_{11}(\frac{1}{3}+\cos^2\theta)-\\
\frac{2}{\sqrt{3}}{\rm Re}\rho_{3-1}\sin^2\theta\cos2\phi-
\frac{2}{\sqrt{3}}{\rm Re}\rho_{31}\sin2\theta\cos\phi],
\end{align*}
where $\rho_{ij}$ are the elements of the production density
matrix. The diagonal elements are real and satisfy
$2(\rho_{33}+\rho_{11})=1$. 
Since the measured distribution in $\phi$ is consistent with being
uniform (this also holds separately for $\cos\theta>0$ and
$\cos\theta<0$ samples), the non-diagonal elements are small.
The result of the fit to the $\cos\theta$ spectrum for the spin
$\frac{3}{2}$ hypothesis is shown in Fig.~\ref{fig3} with a dashed
curve.
The agreement is poor: $\chi^2/d.o.f.=35.1/8$.

The angular distribution for the spin $\frac{5}{2}$ hypothesis 
is~\cite{pilkuhn} 
\begin{align*}
W_{5/2}=\frac{3}{8}
[\rho_{55}2(5\cos^4\theta-2\cos^2\theta+1)+\\
\rho_{33}(-15\cos^4\theta+14\cos^2\theta+1)+
\rho_{11}5(1-\cos^2\theta)^2],
\end{align*}
where non-diagonal elements are ignored. The result of
the fit to the $\cos\theta$ spectrum for the spin $\frac{5}{2}$
hypothesis is shown in Fig.~\ref{fig3} with a solid curve.
The agreement is good: $\chi^2/d.o.f.=12.1/7$.
We find $\rho_{55}=0.09\pm0.02$ and $\rho_{33}=0.00\pm0.03$. 
Thus the $\lcf$ populates mainly the helicity $\pm\frac{1}{2}$ states,
$2\rho_{11}=1-2\rho_{33}-2\rho_{55}=0.82\pm0.05$.

The $\chi^2$ difference of the spin $\frac{1}{2}$ ($\frac{3}{2}$) and
spin $\frac{5}{2}$ fits is distributed as $\chi^2$ with two degrees
(one degree) of freedom, therefore the exclusion level of the spin
$\frac{1}{2}$ ($\frac{3}{2}$) hypothesis is $\sigfo$ ($\sigft$)
standard deviations.

To estimate the systematic uncertainty in the angular analysis of the
$\lcf\to\sigc^{0,++}\pi^{+,-}$ decay we vary the $\lcf$ parameters,
fit interval and background parameterization in the fit to the
$M(\lamc\pip\pim)$ spectrum.
None of the variations alters the exclusion level of the spin
$\frac{1}{2}$ ($\frac{3}{2}$) hypothesis to less than $\sigfosy$
($\sigftsy$) standard deviations.

The Capstick-Isgur quark model predicts the lowest $J^P=\frac{5}{2}^-$
$\lamc$ state at $2900\,\mevm$ and the lowest $J^P=\frac{5}{2}^+$
$\lamc$ state at $2910\,\mevm$~\cite{capstick_isgur}.  The typical
accuracy of quark model predictions is $50\,\mevm$, therefore the
agreement with the experimental value for the $\lcf$ mass is quite
good. The lowest spin $\frac{5}{2}$ states are well separated from the
next $J=\frac{5}{2}$ levels ($3130\,\mevm$ for negative and
$3140\,\mevm$ for positive parities) and from $J=\frac{7}{2}$ levels
($3125\,\mevm$ for negative and $3175\,\mevm$ for positive parities).

Heavy Quark Symmetry predicts 
${\rm R}\equiv\frac{\Gamma(\scst\pi)}{\Gamma(\sigc\pi)}=1.4$ for the
$\frac{5}{2}^-$ state and ${\rm R}=0.23-0.36$ for the
$\frac{5}{2}^+$ state~\cite{hqs,cheng_chua}. 
The measured value ${\rm R}=\ronetwo$ favors the positive parity
assignment for the $\lcf$.

The $\frac{5}{2}^+$ assignment for the $\lcf$ makes it a special state
that lies on the leading $\lamc$ Regge trajectory, whose lower $J^P$
members are the $\frac{1}{2}^+$ $\lamc$ and $\frac{3}{2}^-$
$\Lambda_c(2625)^+$. The $\frac{5}{2}^+$ assignment for the $\lcf$
based on a string model for baryons was proposed in Ref.~\cite{selem}.

In summary, from angular analysis of $\lcf\to\sigc^{0,++}\pi^{+,-}$
decays we find that a $\lcf$ spin hypothesis of $\frac{5}{2}$ is
strongly favored over $\frac{1}{2}$ and $\frac{3}{2}$.
We find first evidence for $\scst\pi$ intermediate states in the
$\lcf\to\lamc\pip\pim$ decays and measure
$\frac{\Gamma(\scst\pi^{\pm})}{\Gamma(\sigc\pi^{\pm})}=\ronetwo$.
This value is in agreement with Heavy Quark Symmetry predictions and
favors the $\frac{5}{2}^+$ over the $\frac{5}{2}^-$ hypothesis for the
spin-parity of the $\lcf$.
We also report the first observation of $\lcs\to\sigc\pi$ decays,
and measure the $\lcf$ and $\lcs$ parameters.

We are grateful to A.~Kaidalov, I.~Klebanov and Yu.~Simonov for
valuable discussions. 
We thank the KEKB group for excellent operation of the
accelerator, the KEK cryogenics group for efficient solenoid
operations, and the KEK computer group and
the NII for valuable computing and Super-SINET network
support.  We acknowledge support from MEXT and JSPS (Japan);
ARC and DEST (Australia); NSFC and KIP of CAS (China); 
DST (India); MOEHRD, KOSEF and KRF (Korea); 
KBN (Poland); MIST (Russia); ARRS (Slovenia); SNSF (Switzerland); 
NSC and MOE (Taiwan); and DOE (USA).

\end{document}